\documentclass[aps,prl,twocolumn,amsmath,amssymb,superscriptaddress]{revtex4}
\usepackage{graphicx}
\usepackage{subfigure}
\usepackage{epsfig}
\usepackage{bm}
\usepackage{ulem}
\usepackage{color}
\usepackage{multirow}
\usepackage{slashed}
\usepackage[colorlinks,
           linkcolor=blue,
           anchorcolor=blue,
           citecolor=blue
           ]{hyperref}
\usepackage{verbatim}

\usepackage{tikz}
\usetikzlibrary{arrows,shapes}
\usetikzlibrary{trees}
\usetikzlibrary{matrix,arrows} 				
\usetikzlibrary{positioning}				
\usetikzlibrary{calc,through}				
\usetikzlibrary{decorations.pathreplacing}  
\usepackage{pgffor}							

\usetikzlibrary{decorations.pathmorphing}	
\usetikzlibrary{decorations.markings}
\tikzset{
    vector/.style={decorate, decoration={snake}, draw},
	provector/.style={decorate, decoration={snake,amplitude=2.5pt}, draw},
	antivector/.style={decorate, decoration={snake,amplitude=-2.5pt}, draw},
    fermion/.style={draw=black, postaction={decorate},
        decoration={markings,mark=at position .55 with {\arrow[draw=black]{>}}}},
    fermiona/.style={draw=red},
    fermionbar/.style={draw=black, postaction={decorate},
        decoration={markings,mark=at position .55 with {\arrow[draw=black]{<}}}},
    fermionnoarrow/.style={draw=black},
    gluon/.style={decorate, draw=black,
        decoration={coil,amplitude=4pt, segment length=5pt}},
    scalar/.style={dashed,draw=black, postaction={decorate},
        decoration={markings,mark=at position .55 with {\arrow[draw=black]{>}}}},
    scalarbar/.style={dashed,draw=black, postaction={decorate},
        decoration={markings,mark=at position .55 with {\arrow[draw=black]{<}}}},
    scalarnoarrow/.style={dashed,draw=black},
    electron/.style={draw=black, postaction={decorate},
        decoration={markings,mark=at position .55 with {\arrow[draw=black]{>}}}},
	bigvector/.style={decorate, decoration={snake,amplitude=4pt}, draw},
}

\usepackage{hyperref}

\def\ben{\begin{equation}}
\def\een{\end{equation}}

\def\be{\begin{equation}}
\def\ee{\end{equation}}
\def\beq{\begin{equation}}
\def\eeq{\end{equation}}
\def\ba{\begin{array}}
\def\ea{\end{array}}

\def\dalemb#1#2{{\vbox{\hrule height .#2pt
       \hbox{\vrule width.#2pt height#1pt \kern#1pt
               \vrule width.#2pt}
       \hrule height.#2pt}}}

\newcommand{\bea}{\begin{eqnarray}}
\newcommand{\eea}{\end{eqnarray}}

\newcommand{\nn}{\nonumber}

\def\Im{{{\frak{Im}}}}
\def\Re{{{\frak{Re}}}}

\def\sech{ {\rm sech}}

\begin{document}

\title{Violation of the viscosity/entropy bound in translationally invariant non-Fermi liquids}
\author{Xian-Hui Ge}
\affiliation{Department of Physics and Shanghai Key Laboratory of
High Temperature Superconductors, Shanghai University, Shanghai, 200444, China}
\author{Shao-Kai Jian}
\affiliation{Institute for Advanced Study, Tsinghua University, Beijing 100084, China}
\affiliation{Condensed Matter Theory Center, Department of Physics, University of Maryland, College Park, Maryland 20742, USA}
\author{Yi-Li Wang}
\affiliation{Department of Physics, Shanghai University, Shanghai, 200444, China}
\author{Zhuo-Yu Xian}
\affiliation{Institute of Theoretical Physics, Chinese Academy of Sciences, Beijing 100190, China}
\author{Hong Yao}
\email{yaohong@tsinghua.edu.cn}
\affiliation{Institute for Advanced Study, Tsinghua University, Beijing 100084, China}
\affiliation{State Key Laboratory of Low Dimensional Quantum Physics, Tsinghua University, Beijing 100084, China}
\affiliation{Department of Physics, Stanford University, Stanford, CA 94305, USA}

\begin{abstract}
Shear viscosity is an important characterization of how a many-body system behaves like a fluid. Here we study the shear viscosity of a strongly-interacting solvable model in two spatial dimensions, consisting of coupled Sachdev-Ye-Kitaev (SYK) islands. As temperature is lowered, the model exhibits a crossover from an incoherent metal with local criticality to a marginal fermi liquid. We find that while the shear viscosity to entropy density ratio satisfies the Kovtun-Son-Starinets (KSS) bound in the marginal Fermi liquid regime, it can strongly violate the KSS bound within a finite and robust temperature range in the incoherent metal regime, implying nearly perfect fluidity of the incoherent metal with local criticality. To the best of our knowledge, it provides the {\it first} translationally invariant example violating the KSS bound with known gauge-gravity correspondence.
\end{abstract}
\date{\today}
\maketitle

\section{I. INTRODUCTION}
Fluid mechanics is among the oldest and the most fundamental subjects in physics.  A generic many-body system with globally conserved quantities, such as mass, energy, and momentum, will exhibit fluidity if the local thermalization time scale is much less than the relaxation time scale of the conserved quantities. As a result, universal properties of a fluid can provide extremely useful insights in understanding correlated many-body systems with complicated interactions between their constitutes, like ultra-cold Fermi gases in the unitary regime and quark-gluon plasma (QGP) produced in relativistic heavy-ion collisions, where no control parameter exists~\cite{Schafer2014}. More recently, owing to the advances of experimental techniques, quantum fluid behaviors are also witnessed in correlated electrons in lattice systems~\cite{Bandurin2016, Crossno2016, Moll2016}. Interestingly, the theory of fluids also receives a boost from the development of holographic principles~\cite{Maldacena1999, Witten1998}. A fundamental characterization of fluids is the shear viscosity that measures the resistance of a fluid to shear stress. Since viscosity generates entropy and causes dissipation, a good fluid should have small shear viscosity. However, the viscosity cannot be arbitrarily small. Namely, like the uncertainty principle, the fundamental laws of nature put a lower bound on the ratio of shear viscosity to entropy. Based on the AdS/CFT correspondence, Kovtun, Son and Starintes conjectured a lower bound (KSS bound) on the ratio of shear viscosity to entropy in strongly coupled non-quasiparticle systems~\cite{KSS2005}, i.e., $\eta/\mathcal{S} \ge 1/4\pi$, where $\eta$ and $\mathcal{S}$ refer to shear viscosity and entropy density, respectively.

The closer the ratio, $\eta/\mathcal{S}$, of a many-body system is to the KSS bound, the better it behaves as a perfect fluid. Thus, it is of great interest and importance to explore scarce examples that saturate, or even violate the bound. Among holographic systems, the KSS bound is obeyed in Einstein gravity with both rotational and translational symmetries, while a weaker bound~\cite{liu08,Kats2009, cai,sera,ge, Myers2009, Camanho2010} is obeyed in higher-derivative gravity theory. When rotational symmetry is broken, like the anisotropic black branes~\cite{mateos,rebhan,gl}, certain component of shear viscosity tensor may violate the KSS bound in a parametric manner which was recently illustrated in an anisotropic Dirac fluid~\cite{Schmalian2018}. Moreover, the black brane solution for Gauss-Bonnet massive gravity and Rastall AdS massive gravity show violation of KSS bound~\cite{bb}. For isotropic black branes with linear axion fields, the KSS bound can also be violated; but shear viscosity does not have a hydrodynamic interpretation since momentum is no longer conserved~\cite{RA,hartnoll,matteo,poov,pope,ling,wang}.

For many-body systems, the minimal of the ratio $\eta/\mathcal{S}$ normally occurs at the fixed point exhibiting emergent conformal symmetry, where the quasiparticle description often invalidates. When the fixed point locates at zero temperature, the ratio should be a universal number associated with the universality class of the fixed point. Such examples include the electron fluid in graphene~\cite{Schmalian2009}, the Luttinger-Abrikosov-Beneslavskii phase in three dimensional quadratic band touching semimetal~\cite{Dumitrescu2015}, and Ising nematic quantum critical point in 2D metals~\cite{Patel2016}. However, if the fixed point locates at finite temperature, the ratio shows a non-universal behavior as a function of temperature. The well-studied unitary quantum gases and the QGP fall into this class \cite{Teaney2003,Csernai2006,cao2011,Enss2011,Bruun2007}. In unitary quantum gases, the minimal of the ratio occurs at an intermediate temperature range associated with the superfluid transition, providing possible examples violating the KSS bound~\cite{Pakhira2015}, while at the zero-temperature limit the gapless Goldstone modes lead to a divergent ratio.

\begin{figure}[t]
\subfigure[]{\label{sykham}
\includegraphics[height=3.4cm]{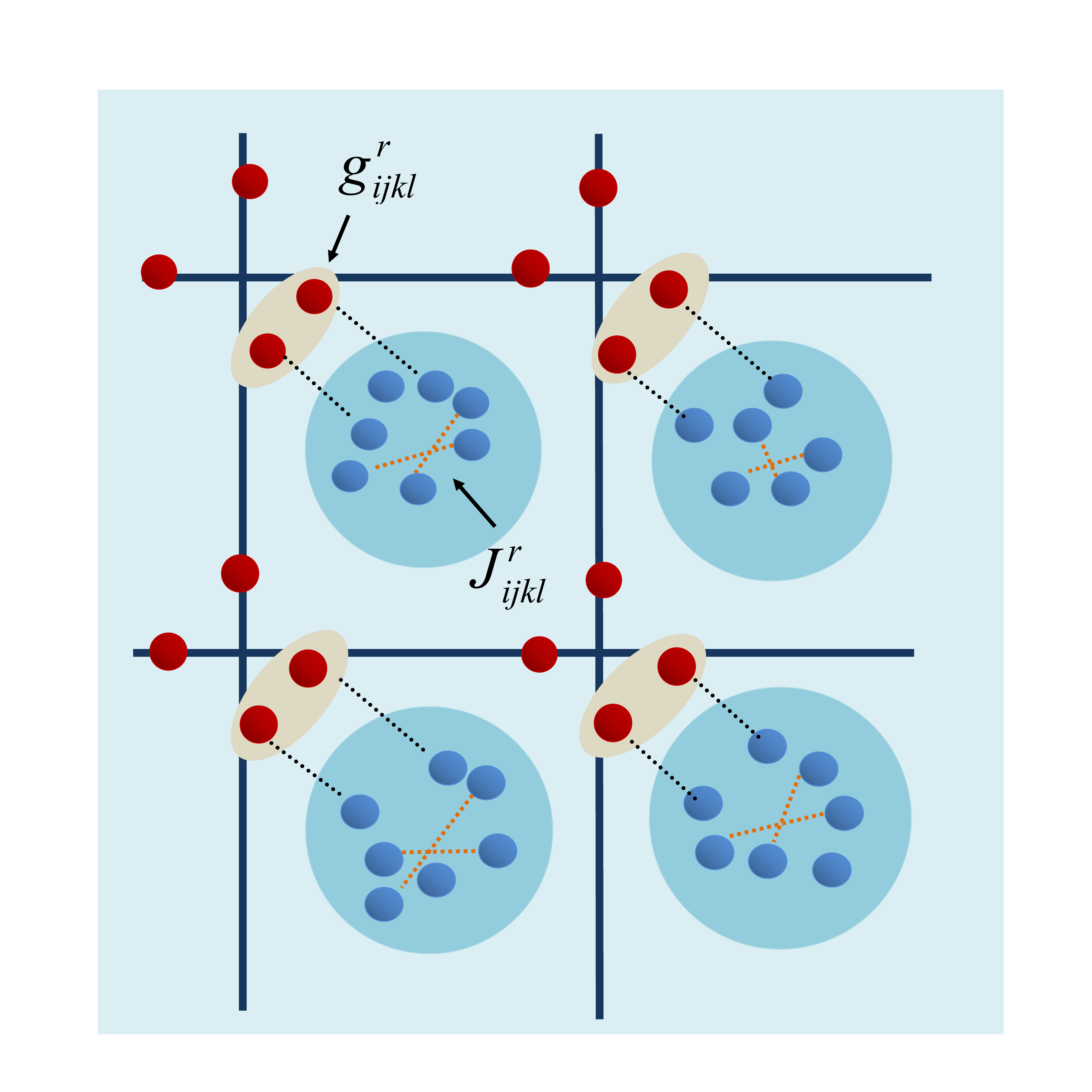}}\quad\quad
\subfigure[]{\label{ratio}
\includegraphics[height=3.4cm]{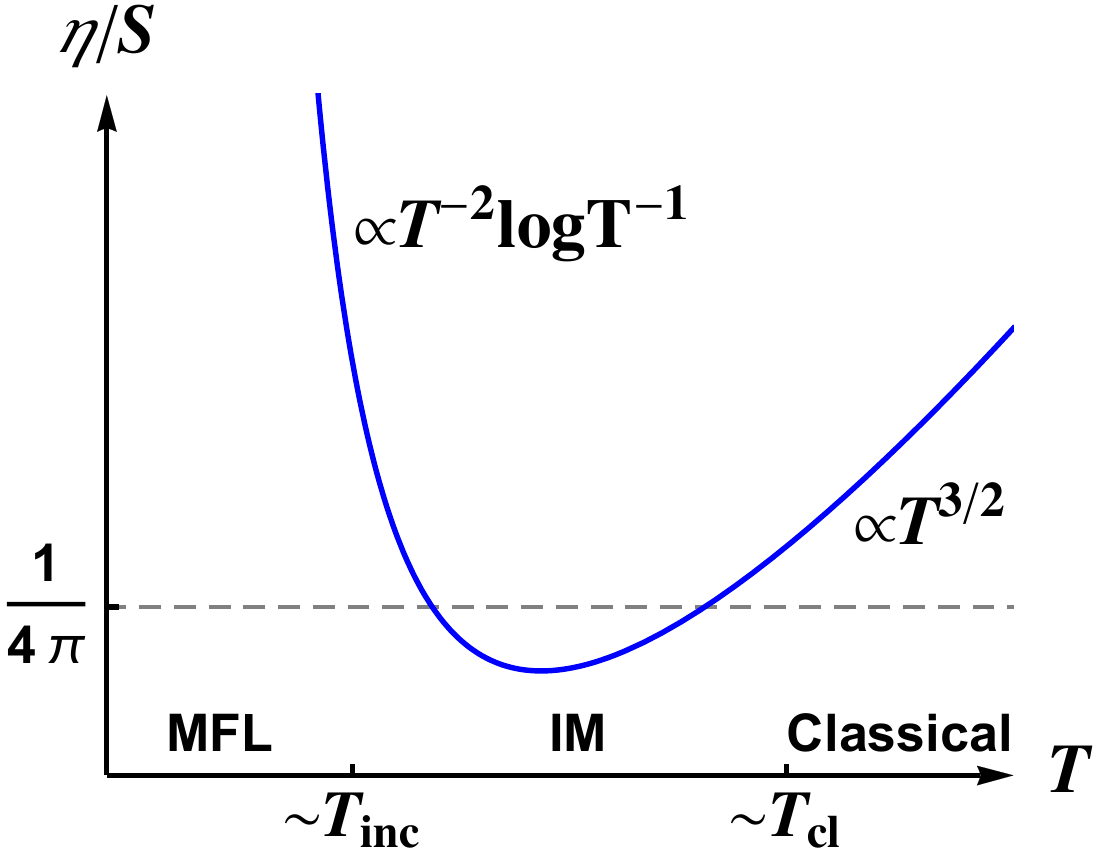}}
\caption{(a) A cartoon of the model. The red and blue dots represent the conduction electrons ($c$ fermions), and ($f$ fermions), respectively. The black dotted lines and orange dashed lines indicate the interactions between between $f$ and $c$ fermions and   self-interaction of $f$ fermions, respectively. (b) A schematic plot of the ratio $\eta/\mathcal{S}$ as a function of temperature. There are three regimes, marginal Fermi liquid (MFL), IM (incoherent) and semi-classical regime, exhibiting different behaviors. The ratio violates the KSS bound indicated by the dashed line in the IM regime.}
\vspace{-0.25cm}
\end{figure}

Recently, Patel {\it et al.}~\cite{Patel2018} and Chowdhury {\it et al.}~\cite{Senthil2018} constructed a 2D strongly correlated solvable model, consisting of coupled Sachdev-Ye-Kitaev (SYK) islands as shown in Fig.~\ref{sykham}. This model is of great interest due to the fact that the SYK model is believed to have a gravity dual~\cite{Kitaev2015a, Kitaev2015b, Sachdev1993, Polchinski2016, Maldacena2016, Jensen2016, note2} with maximal chaos~\cite{chaosbound}, and that though the model exhibits marginal Fermi liquid (MFL) with well-defined quasiparticle at low temperature, it exhibits an intermediate-temperature incoherent metal (IM) regime where the quasiparticle description invalidates.
Here, we consider a translationally invariant version of such model~\cite{Senthil2018}, and evaluate the shear viscosity by using the Kubo formula at large-$N$ limit. As indicated in Fig.~\ref{ratio}, in the MFL regime with $T<T_{\text{inc}}$, we find $\eta/\mathcal{S} \propto T^{-2}$; the ratio obeys a KSS-like bound and diverges at zero-temperature limit. For $T>T_{\text{cl}}$ where the system can be treated classically, we have $\eta/\mathcal{S} \propto T^{3/2}$~\cite{note1}. Thus, the ratio necessarily exhibits a minimal in the intermediate temperature. Interestingly, the ratio can strongly violate the KSS bound in a robust temperature range of the IM regime, not only implying a nearly perfect fluidity of the coupled local critical SYK models, but also providing the {\it first} translationally and rotationally invariant example violating the KSS bound with known gauge-gravity correspondence.

\section{II. THE MODEL}
We consider a 2D lattice model with $M$ flavors of conduction fermions $c_{ri}$, $i$=$1,\!\cdots\!,M$, and $N$ flavors of valence fermions $f_{rj}$, $j$=$1,\!\cdots,\!N$, on each site $r$, as shown in Fig.~\ref{sykham}:
\begin{eqnarray}\label{Hamiltonian}
\centering
H &=& -\sum_{rr'} \sum_{i=1}^{M}( t_{rr'} c_{ri}^{\dagger}c_{r'i}+h.c.) + \sum_r \Big[ - \mu_c \sum_{i=1}^{M}c_{ri}^{\dagger}c_{ri}  \nn\\
 &&-\mu_f \sum_{i=1}^{N}f_{ri}^{\dagger}f_{ri} +\sum_{i,j=1}^{N}\sum_{k,l=1}^{M} \frac{g_{ijkl}}{NM^{1/2}} f_{ri}^{\dagger}f_{rj}c_{rk}^{\dagger}c_{rl} \nn \\
 && +\sum_{i,j,k,l=1}^{N} \frac{J_{ijkl}}{N^{3/2}}f_{ri}^{\dagger}
f_{rj}^{\dagger}f_{rk}f_{rl} \Big] .
\end{eqnarray}
where $t_{rr'}$ is the hopping amplitude of $c$ fermions between sites $r$ and $r'$, and $\mu_i$ ($i=c,f$) denote the chemical potential of $c$ and $f$ fermions, respectively.
The local interaction strength $g_{ijkl}$ and $J_{ijkl}$ are random numbers which satisfies
$ \langle\langle J_{ijkl}J_{lkij}\rangle \rangle =\frac{J^2}{8}$ and
$\langle\langle g_{ijkl}g_{lkij}\rangle \rangle=g^2$ and all other $\langle\langle...\rangle \rangle$ are vanishing. Here $\langle\langle ...\rangle \rangle$ means disorder-average.
Note that the coupling constants $g_{ijkl}$ and $J_{ijkl}$ on different sites not only have the same distribution, but are  identical in each realization. In the following, we choose the hopping amplitude to be a function depending on $|r-r'|$, for instance, $t_{rr'}= t \delta_{r',r+\hat e_i}$, where $\hat e_i$ is the primitive lattice vector. As a result, the Hamiltonian is translationally invariant.
If $g=0$, the model can be viewed as two independent subsystems: the conducting $c$ fermions with a hopping $t_{rr'}$, and the local $f$ fermions with SYK interaction at each site. Finite $g>0$ will couple the two subsystems, as illustrated in Fig.~\ref{sykham}. They interact through a random exchange with effective strength $g$,
similar to the Kondo lattice model ~\cite{s19,s20,s21}.

We consider large $N$ and $M$ limit, while keep their ratio, $M/N$, fixed. The Green's functions are given by~\cite{Senthil2018}, $ G^{c}(\textbf{k},i\omega)=[i\omega_n-\epsilon_k+\mu_c-\Sigma_{cf}(\textbf{k},i\omega_n)]^{-1}$ and $G^{f}(\textbf{k},i\omega_n)=[i\omega_n+\mu_f-\Sigma'_{cf}(\textbf{k},i\omega_n)-\Sigma_{f}(\textbf{k},i\omega_n)]^{-1}$,
where $\textbf k$ and $\omega_n$ denote momentum and Matsubara frequency, $\Sigma_{cf}, \Sigma_{cf}'$ and $\Sigma_{f}$ refer to self-energies from the coupling between $c$ and $f$ fermions and self-interaction of $f$ fermions, respectively. Local critical $f$ fermion propagator, i.e., $G^f(\textbf{k}, i \omega_n) = G^f(i \omega_n)$, is always a consistent solution to saddle point equations~\cite{supp}. Especially, in the limit $M/N=0$, the saddle point equations of $f$ fermions are identical to the zero-dimensional complex SYK model with the following conformal-limit solutions~\cite{Sachdev2015}
\bea
G^{f}(\tau)=-\frac{\pi^{\frac{1}{4}}\cosh^{\frac{1}{4}}(2\pi \mathcal{E})}{J^{\frac{1}{2}}\sqrt{1+e^{-4\pi \mathcal{E}}}}\bigg(\frac{T}{\sin(\pi T \tau)}\bigg)^{\frac{1}{2}}e^{-2\pi \mathcal{E}T\tau },\nonumber
\eea
where $\mathcal{E}$ is a parameter controlling the particle-hole asymmetry, and $ \tau \in[0, \beta]$ is the imaginary time.

Now, moving to the propagator of $c$ fermion, we will follow Ref.~\cite{Patel2018} closely. Though the model in Ref.~\cite{Patel2018} breaks translational symmetry by the locally independent disorder, we show in appendix that at $\frac{M}N \ll 1$, both models have the same saddle point solutions. The translational symmetry and the resulting momentum conservation equation are also shown in appendix. In the limit $g^2 \gg t J$, there exists a crossover temperature, $T_{\rm inc} \sim \frac{t^2 J}{g^2} $, between the MFL regime in the lower temperature and the IM regime in higher temperature. When $T \ll T_{\rm inc}$, the hopping term between conduction electrons dominates, and the self-energy of the $c$ fermion yields~\cite{Patel2018, supp}
\begin{eqnarray}\label{selfMFL}
\Sigma_{cf}^{\rm MFL}(i\omega_n)&=&\frac{ig^2 T}{2J t \cosh^{1/2}(2\pi \mathcal{E})\pi^{3/2}} \bigg(\frac{\omega_n}{T}\ln\bigg(\frac{2\pi T e^{\gamma_{E}-1}}{J}\bigg)\nonumber\\
&&+\frac{\omega_n}{T}\psi\left(-\frac{i \omega_n }{2 \pi  T}\right)+\pi\bigg),
\end{eqnarray}
where $\psi$ is the digamma function, and $\gamma_E=0.577$ is the Euler-Mascheroni constant. The self-energy shows that the $c$ fermions exhibit a MFL behavior. Indeed, in this regime, the model a linear-in-$T$ resistivity as well as a $T\ln T$ entropy density~\cite{Patel2018,Moca1996}, i.e., $\mathcal{S}_{\rm MFL}\sim  \frac{g^2 M}{ J t^{2}}(T+T\ln \frac{J}T)$.

On the other hand, when $T>T_{\rm inc}$, the interacting term between the conduction and the valence band electrons dominates. Since the interacting term is local, the $c$ fermion propagator will also exhibit local critical behavior~\cite{Patel2018, supp}. The $c$ fermion self-energy reads~\cite{Patel2018, supp}
\begin{eqnarray}\label{selfM}
\Sigma_{cf}^{\rm IM}(i\omega_n)&=&\frac{iT^{\frac{1}{2}}g^2 \Lambda^{\frac{1}{2}} \nu^{\frac{1}{2}}(0)(-1)^{\frac{1}{4}}(1+e^{4\pi \mathcal{E}_c })^{\frac{1}{2}}e^{2\pi\mathcal{E}}}{\pi^{\frac{1}{4}}J^{\frac{1}{2}}2^{\frac{3}{2}}(i+e^{2\pi\mathcal{E}_c})\cosh^{\frac{1}{4}}(2\pi\mathcal{E})}\nonumber\\
&&\times \frac{\Gamma(\frac{3}{4}+i\mathcal{E}_c+\frac{\omega_n}{2\pi T})}{\Gamma(\frac{1}{4}+i\mathcal{E}_c+\frac{\omega_n}{2\pi T})},
\end{eqnarray}
where $\Gamma$ denotes Gamma function, and $\mathcal{E}_c$ is a parameters related to the conduction band filling. At small $\mu_f/J$,  $\mu_c/g$ limit, $ \mathcal{E}\simeq -\frac{\mu_f/J}{\pi^{1/4}\sqrt{2}}$ and $\mathcal{E}_c\simeq -\pi^{1/4}\cosh^{1/4}(2\pi \mathcal{E})\mu_c/g$~\cite{Patel2018}.
The form of self-energy indicates the quasiparticle does not exist, and the conduction electrons enter the IM regime. As the Green's functions of both $c$ and $f$ fermions are local SYK-type~\cite{Patel2018}, the entropy density scales as $\mathcal{S}_{\rm IM} \sim M \frac{J T}{g^2} + N \frac{T}J$, where the first and second term come from $c$ fermions and $f$ fermions, respectively \cite{Patel2018}.

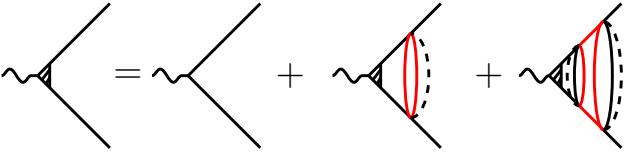
\begin{figure}
\begin{center}
\begin{tikzpicture}[line width=1.1 pt, scale=0.8]
	\draw(0,0)--(1.2,1.2);
	\draw(0,0)--(1.2,-1.2);
	\draw[vector] (180:0.6)--(0,0);
	\draw (0.2,0.2)--(0.2,-0.2);
	\draw (0.07,-0.07)--(0.2,0.1);
    \draw (0.14,-0.14)--(0.2,-0.03);

    \node at (-0:1.5) {\Large{=}};

\begin{scope}[shift={(2.5,0)}]
	\draw(0,0)--(1.2,1.2);
	\draw(0,0)--(1.2,-1.2);
	\draw[vector] (180:0.6)--(0,0);
    \end{scope}

    \node at (4.2,0) {\Large{+}};
    \begin{scope}[shift={(5.5,0)}]
    \draw(0,0)--(1.2,1.2);
	\draw(0,0)--(1.2,-1.2);
	\draw[vector] (180:0.6)--(0,0);
	\draw (0.2,0.2)--(0.2,-0.2);
	\draw (0.07,-0.07)--(0.2,0.1);
    \draw (0.14,-0.14)--(0.2,-0.03);
    \draw[fermiona](0.7,0.7)arc (-270:-90:0.1 and 0.7);
    \draw[fermiona](0.7,0.7)arc (90:-90:0.1 and 0.7);
    \draw[scalarnoarrow](0.7,-0.7) arc (-90:90:0.3 and 0.7);

     \end{scope}

 \begin{scope}[shift={(8.5,0)}]
	\node at (-1,0) {\Large{+}};	
    \draw[vector] (-0.5,0)--(0,0);
	\draw (0,0)--(0.5,0.5);
	\draw (0,0)--(0.5,-0.5);
    \draw (0.9,0.9)--(1.2,1.2);
	\draw (0.9,-0.9)--(1.2,-1.2);
    \draw [fermiona] (0.5,0.5)--(0.9,0.9);
	\draw [fermiona] (0.5,-0.5)--(0.9,-0.9);
    \draw (0.2,0.2)--(0.2,-0.2);
	\draw (0.07,-0.07)--(0.2,0.1);
    \draw (0.14,-0.14)--(0.2,-0.03);
    \draw (0.5,0.5) arc (-270:-90:0.08 and 0.5);
    \draw[fermiona] (0.5,0.5) arc (-270:-90:-0.08 and 0.5);
    \draw[scalarnoarrow](0.5,0.5) arc (-270:-90:0.2 and 0.5);
     \draw[scalarnoarrow](0.9,-0.9) arc (-90:90:0.3 and 0.9);
    \draw (0.9,-0.9) arc (-90:90:0.15 and 0.9);
     \draw[fermiona] (0.9,-0.9) arc (-90:90:-0.15 and 0.9);
    \end{scope}
 \end{tikzpicture}
 \caption{The ladder diagram shows the  self-consistent equation for shear viscosity vertex. The black and red solid lines represent the Green's function of $c$ fermions and $f$ fermions, respectively. The dashed line represents disorder average and the shaded vertex represents full vertex.}\label{vertex1}
\end{center}
\end{figure}

\section{III. SHEAR VISCOSITY}
The shear viscosity is usually evaluated via the Kubo formula $ \eta =\lim_{\omega\rightarrow 0}\frac{1}{\omega}{\rm Im} G_{R}^{xy,xy}(\omega,0)$, where $G_{R}^{xy,xy}$ is the retarded Green's function of $xy$ component of the energy-momentum tensor, i.e.,
\bea
i G_{R}^{xy,xy}(\omega,\textbf{p})=\int dt d\textbf{x}e^{i (\omega t-\textbf{p} \cdot \textbf{x})}  \theta(t)\langle[T_{xy}(t,\textbf{x}),T_{xy}(0,0)]\rangle. \nn
\eea
where $\theta(t)$ denotes the step function such that $\theta(t)=1$ for $t \ge 0$ and zero otherwise, and $[...]$ is commutator. In the following, we consider the isotropic dispersion $\epsilon_{\bf k}=\frac{{\bf k}^2}{2m}-\frac{\Lambda}{2}$ with $-\Lambda/2\le\epsilon\le \Lambda/2$. Generalization to other dispersions is straightforward, and won't change our results qualitatively. Note that the lattice constant has been taken to be $1$, so momentum $\bf k$ becomes dimensionless, and we have the relations $m \sim\frac{1}{t} \sim\frac{1}{\Lambda} \sim \nu(0)$, where $\nu(0)$ denotes the density of states at Fermi level. For the isotropic dispersion, the density of state is a constant, $\nu(\epsilon) \equiv \int_{\bf k} 2\pi \delta(\epsilon - \epsilon_{\bf k}) = \nu(0)$, $\int_{\bf k} \equiv \int \frac{d^2 \bf k}{(2\pi)^2}$, irrespective of the energy.   The tensor $T_{xy}$ of $c$ fermions is given by $T_{xy}({\bf p})=\sum_i \int_{\bf k} c^{\dagger}_{{\bf k}i} \Gamma_0({\bf p;k}) c_{{\bf k+p},i}+{\rm H.c.} $, where $c_{{\bf k}i}=\int d{\bf x} c_{{\bf x}i} e^{i{\bf k\cdot x}}$, and $\Gamma_0({\bf p;k}) = \frac{(k_x+\frac{p_x}2) (k_y+\frac{p_y}2) }{m}$ for the isotropic dispersion.

\begin{figure}
\begin{center}
\begin{tikzpicture}[line width=1.5 pt, scale=1]
    \draw[vector](1.25,0)--(2,0);
    \draw[vector](3,0)--(3.75,0);
    \draw[fill=black](2.5,0) circle (0.5);
    \draw[fill=white](2.5,0) circle (0.5);
    \draw(2.2,-0.35) arc (-90:90: 0.15 and 0.38);
    \node at (4.25,0) {\Large{=}};
    \begin{scope}[shift={(4.5,0)}]
    \draw[vector](0.25,0)--(1,0);
    \draw[vector](2,0)--(2.75,0);
    \draw[fill=black](1.5,0) circle (0.5);
    \draw[fill=white](1.5,0) circle (0.5);
    \end{scope}
    \node at (7.75,0) {\Large$+$};
    \node at (8.9,0) {\Large$\mathcal{O}\left(\frac1{N^2}\right)$};
\begin{scope}[shift={(2.2,0)}]

	\clip (0,0) ellipse (0.15 and 0.40);
\foreach \x in {-0.9,-0.8,...,0.9}
\draw [line width=1 pt](\x,-0.6)--(\x+0.6,0.6);
\end{scope}
\end{tikzpicture}
\caption{The Feynman diagram for the calculation of $\langle T_{xy}T_{xy}\rangle$ at leading order in $1/N$, where the vertex correction vanishes. The black lines represent the Green's function of $c$ fermions.}\label{FD2}
\end{center}

\end{figure}
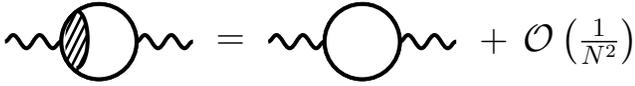

As shown in Fig.~\ref{vertex1}, to the leading nontrivial order in large-$N$ limit, the self-consistent equation for the full vertex $\Gamma$ is
\bea
\Gamma(p;q)&=&\Gamma_{0}({\bf p;q})+ \frac1{N} \sum_i \int_{q'} \mathcal{F}^{(i)}(p; q,q') \Gamma(p;q'),
\eea
where $\int_k \equiv \int_{k_0} \int_{\bf k}$, $\int_{k_0} \equiv T\sum_{\omega_n}$ and $F^{(i)}$ is represented in the second and third diagram in Fig.~\ref{vertex1}, i.e.,
\bea
	\mathcal{F}^{(1)} = -g^2 \int_k G^f(q-q'+k)G^f(q'-q-k)G^c(q')G^c(p+q').\nn
\eea
Because we are interested in the uniform case, i.e., ${\bf p=0}$, 
\bea\label{F1}
&&\int_{q'}\mathcal{F}^{(1)}({\bf 0}, p_0 ;q,q') \Gamma({\bf 0}, p_0;q') \nn \\
&& \quad\quad = -g^2\int_{k,q'_0}G^f(q-q'+k)G^f(q'-q-k) \nn\\
&& \quad\quad \times \int_{{\bf q'}} G^c({\bf q'},q_0')G^c({\bf q}',p_0+q_0')\Gamma({\bf 0}, p_0;q'),
\eea
Eq.~\ref{F1} vanishes since it is odd in $q'_x$ (or $q'_y$). Owing to the same reason, we find that $\mathcal{F}^{(2)}$ on the right-hand side in Fig.~\ref{vertex1} also vanishes. Therefore, the vertex corrections vanish, $\Gamma({\bf 0}, p_0;q) =\Gamma_0({\bf 0;q})=\frac{q_x q_y}{m}$. Thus, to leading order in $1/N$, the shear viscosity is given by the sum over the set of ladder diagrams shown in Fig. \ref{FD2}, and the spectral representation of shear viscosity is~\cite{supp}
\begin{equation}\label{eta}
\eta=\frac{M}{4\pi}  \int_{-\infty}^{+\infty}d\omega \left(-\frac{\partial n_F(\omega)}{\partial \omega}\right)
\int_{-\infty}^{+\infty}d\epsilon \Theta_{xy}(\epsilon)A^c(\omega,\epsilon)^2,
\end{equation}
where $n_F(\omega)=1/(e^{\beta\omega}+1)$ is the Fermi-Dirac distribution, $ A^c(\omega,\epsilon)=-2 {\rm Im} [G^c(i\omega_n\to \omega+ i 0^+, \epsilon)]$ denotes the spectral function, and $ \Theta_{xy}(\epsilon)=\int \frac{d^2 \bf k}{(2\pi)^2} (\frac{k_x k_y}{m})^2 \delta(\epsilon-\epsilon_{\bf k})$ is the \textit{transport density of states} for shear viscosity.

\section{IV. SHEAR VISCOSITY IN MFL REGIME}
In the MFL regime, the Fermi surface is well defined and the leading temperature-dependence contribution to viscosity comes from the states near Fermi surface, $\epsilon=0$. This allows us to approximate $\Theta_{xy}(\epsilon)$ by the value at Fermi surface, i.e., $\Theta_{xy}(\epsilon) \approx \frac{k_F^4}{16\pi m^2} \nu(0)$, and to extend the range of the integral of $\epsilon$ to infinity~\cite{supp}. Finally, we have
\begin{eqnarray}\label{etaMFL}
\eta_{\rm MFL}(T) &=&\frac{M \nu(0)}{64 m^2 T} \int_{-\infty}^{\infty} \frac{d\omega}{2\pi} \text{sech}^2(\frac{\omega}{2T}) \frac1{|\Im\Sigma_{cf}^{\rm MFL}(\omega)|} \nn \\
&\approx&0.0300627\frac{Mt^2J}{g^2T}\cosh^{\frac12}(2\pi\mathcal{E}).
\end{eqnarray}
Dividing the viscosity by the entropy density contributed by $c$ fermions, $\mathcal{S}^{\rm MFL}_c\sim \frac{g^2 M}{ J t^{2}}T\ln \frac{J}T$, the shear viscosity to entropy ratio at low temperature scales as
\begin{equation}
\frac{\eta_{\rm MFL}}{\mathcal{S}^{\rm MFL}_c}
\sim \cosh^{\frac12}(2\pi\mathcal{E})\frac{J^2  t^4 }{  g^4 T^2 \ln(\frac{J}{T})}.
\end{equation}
Since $T\ll T_{\rm inc}$ in MFL regime, the ratio is larger than a constant, $ \eta_{\rm MFL}/\mathcal{S}_c^{\rm MFL} \gg 1/\ln(J/T_{\rm inc}) = 1/2\ln(g/t)$. At zero temperature limit, $\eta_{\rm MFL}/\mathcal{S}_c^{\rm MFL}$ diverges, as shown in Fig.~\ref{ratio}.

For the system with (marginally) well-defined quasiparticle, the shear viscosity is actually proportional to the lifetime of quasiparticle, as indicated in Eq.~(\ref{etaMFL}). The quasiparticle lifetime in the MFL is $\tau \propto T^{-1}$, which leads the scaling form of shear viscosity $\eta \propto T^{-1}$ (up to logarithmic corrections). Note that for Fermi liquid, the quasiparticle lifetime, $\propto T^{-2}$, leads to the well-known result $\eta \propto T^{-2}$. More concretely, the inverse lifetime of $c$ fermions in the MFL regime is~\cite{Patel2018}
\begin{equation}\label{gamma}
\gamma=\frac{g^2 T}{\pi^{\frac12}t J \cosh^{\frac12}(2\pi\mathcal{E})}.
\end{equation}
Then we can estimate the viscosity to be
\begin{equation}
\eta_{\rm MFL} \approx \varepsilon \gamma^{-1}\sim \frac{Mt^2J}{g^2T}\cosh^{\frac12}(2\pi\mathcal{E}),
\end{equation}
where $\varepsilon$ is the energy density which scales as $\varepsilon\sim M t$, agreeing with the result in Eq.~(\ref{etaMFL}).

\section{V. SHEAR VISCOSITY IN IM REGIME}
In the IM regime, the $c$ fermions exhibit local critical behavior, and there is no notion of Fermi surface. Thus, in contrast to the case of MFL, we should calculate $\Theta_{xy}(\epsilon)$ in the full spectrum instead of approximating it at the fermi surface~\cite{supp}, $\Theta_{xy}(\epsilon)= \frac m{4\pi} \left(\epsilon+\frac\Lambda2 \right)^2 \theta\left(\frac\Lambda2-|\epsilon|\right)$.
A technical advantage occurs owing to the local critical form of $c$ fermions' propagator in the IM regime, namely, the spectral function is independent of $\epsilon$, $A^c_{\rm IM}(\omega, \epsilon) = A_{\rm IM}^c(\omega)$. As a result, the shear viscosity splits into two independent integrations,
\bea\label{etaIM}
\eta_{\rm IM}= \frac{M }{16 \pi T} \int d \epsilon \Theta_{xy}(\epsilon) \int d\omega \sech^2(\frac{\omega}{2T}) A^c_{\rm IM}(\omega)^2,
\eea
both of which can be evaluated directly~\cite{supp}, and the final result is
\bea\label{etaIM}
\eta_{\rm IM}(T)=\frac{M\pi^{\frac12}}{24} \frac{\Lambda^2 J}{g^2 T}  \frac{\cosh^{\frac12}(2\pi \mathcal{E})}{\cosh(2\pi \mathcal{E}_c)}.
\eea
In the IM regime, the entropy density corresponds to $c$ fermions is given by $\mathcal{S}^{\rm IM}_c \sim M \frac{JT}{g^2}$, so the ratio between shear viscosity and entropy density is given by
\begin{equation}
\frac{\eta_{\rm IM}}{\mathcal{S}_c^{\rm IM}} \sim \frac{\cosh^{\frac12}(2\pi\mathcal{E})}{\cosh(2\pi\mathcal{E}_c)}\frac{\Lambda^2}{T^2} .
\end{equation}
If $\Lambda \ll J$, there exists a robust temperature window in the IM regime, i.e., $\Lambda \ll T\ll {\rm min}(J, g^2/J)$, such that the KSS bound is strongly violated!

In fact, the scaling form of the shear viscosity obtained in the IM regime, $\eta \propto T^{-1}$, is a universal property for local critical systems. In local critical regime, the local interaction dominates over hoppings, and in turn dictates the scaling dimension of fermions. The most generic local interaction allowed by $U(1)$ symmetry is of quartic order. Thus, the local critical freedoms, i.e., the $c$ fermions in our case, have scaling dimension 1/4, and consequently the spectral weight $A \propto T^{-1/2}$. Furthermore, the local criticality also renders the vertex correction vanishing, and leads to the spectral representation of shear viscosity, as shown in Eq.~(\ref{eta}). These reasons lead to the scaling form of shear viscosity $\eta \propto T^{-1}$. Note that though the scaling form is the same in the MFL regime, the origins behind them are different, i.e., the shear viscosity is determined by quasiparticle lifetime in the MFL as discussed before. The essential point for the violation of the KSS bound is that the scaling form in the IM regime can survive in an intermediate-temperature range, which lead to a robust energy window violating the bound, as indicated in Fig.~\ref{ratio}. In the discrete translationally symmetric system considered here, the only process that can relax the momentum is electron-electron umklapp scattering. However, the $c$-fermion density can be tuned small enough to suppress the umklapp process in low energy and long-distance,
so that the system is essentially momentum-preserving and hydrodynamics emerges in both the MFL and the IM regimes \cite{lucas2017}. It calls for further experiments to establish whether or not the electron fluids in strange metals are in the hydrodynamic regime

\section{VI. DISCUSSION AND CONCLUSIONS}
Though a similar violation of the KSS bound is also reported in unitary quantum gases by dynamic mean field theory calculation~\cite{Pakhira2015}, the SYK model has a  better holographic interpretation~\cite{Kitaev2015b, Maldacena2016} and analytical controllability than the model used in Ref.~\cite{Pakhira2015}. Thus our calculations provide the {\it first} translationally invariant example violating the KSS bound with known gauge-gravity correspondence. Moreover, as indicated in Ref.~\cite{Senthil2018, Xian2018}, we also expect that the model in this paper has a description of semi-holography: $f$ fermions form the bulk geometry while $c$ fermions live on the boundary. From this point of view, the $\eta/\mathcal{S}_c$ we calculate here is different from the one calculated in those full-holographic models, where the entropy is black hole entropy. To compare our result with those full-holographic results, one should replace the $\mathcal{S}_c$ in $\eta/\mathcal{S}_c$ by the entropy density of the whole system consisted of both $f$ fermions and $c$ fermions. Since $\mathcal{S}_f \propto N \gg \mathcal{S}_c \propto M$, we have $\eta/\mathcal{S}_f \propto M/N \rightarrow 0$, at the $M/N\ll 1$ limit. Here, the KSS bound is violated trivially, since the entropy density comes from an immobile contribution, $S_f$, with $U(1)$ symmetry at each site.

In conclusion, we investigated the shear viscosity in a translationally invariant, strongly correlated solvable model~\cite{Patel2018, Senthil2018}. By using Kubo formula, we obtained the interesting behaviors of shear viscosity as a function of temperature. In the MFL regimes, the shear viscosity is related to the quasiparticle lifetime; in the IM regimes, the result is more general and can be inferred from local criticality. As shown in Fig.~\ref{ratio}, we further find an interesting robust temperature range in the IM regime where the ratio of shear viscosity to entropy density, $\eta/\mathcal{S}$, can strongly violate the KSS bound. To the best of our knowledge, it is for the {\it first} time that the perfect fluidity behaviors are discovered in the coupled local critical SYK models in an intermediate-temperature range. We believe that our results could shed new light to understanding shear viscosity of strongly correlated systems.

\section{ACKNOWLEDGEMENT} 
We would like to thank Wei-Jia Li, Hong L$\ddot{u}$, Sang-Jin Sin, Yu Tian and Shao-Feng Wu for helpful discussions. This work is partly supported by NSFC (No.11875184 $\&$ No.11805117), NSFC under Grant No. 11825404 (S.-K.J. and H.Y.), the Simons Foundation via the It From Qubit Collaboration (S.-K.J.), and NSFC under Grant No.~11575195 (Z.-Y.X.). Z.-Y.X. is also supported by the National Postdoctoral Program for Innovative Talents BX20180318. X.-H.G. would also like to thank Hanyang University for the hospitality during the APCTP focus program ``Holography and Geometry of Quantum Entanglement". 

\section{APPENDIX A: SADDLE POINT SOLUTIONS}
\renewcommand{\theequation}{A\arabic{equation}}
\setcounter{equation}{0}
\renewcommand{\thefigure}{A\arabic{figure}}
\setcounter{figure}{0}
\renewcommand{\thetable}{A\arabic{table}}
\setcounter{table}{0}

Summing the relevant Feynman diagrams in the large-$N$ limit~\cite{Senthil2018}, the saddle-point equations are given by
\bea
G^{c}(\textbf{k},i\omega)&=&\frac{1}{i\omega_n-\epsilon_k+\mu_c-\Sigma_{cf}(\textbf{k},i\omega_n)},\\
G^{f}(\textbf{k},i\omega_n)&=&\frac{1}{i\omega_n+\mu-\Sigma'_{cf}(\textbf{k},i\omega_n)-\Sigma_{f}(\textbf{k},i\omega_n)}, \\
\Sigma_{cf}(\textbf{k},i\omega_n) &=& -g^2\int_{k'}G^{c}(\textbf{k}',i\omega_{n'})\notag\\ &&\Pi_f(\textbf{k}+\textbf{k}',i \omega_n+i\omega_{n'}), \label{Sigma_cf}\\
\Sigma'_{cf}(\textbf{k},i\omega_n)&=& -\frac{M}{N}g^2\int_{k'}G^{f}(\textbf{k}',i\omega_{n'}) \notag\\
&&\Pi_c(\textbf{k}+\textbf{k}',i \omega_n+i\omega_{n'}), \\
\Sigma_{f}(\textbf{k},i\omega_n) &=& -J^2\int_{k'} G^{f}(\textbf{k}',i\omega_{n'})\notag\\ &&\Pi_f(\textbf{k}+\textbf{k}',i \omega_n+i\omega_{n'}),\label{Isix} \\
 \Pi_f(\textbf{q}, i\Omega_n)&=&\int_{k}G^{f}(\textbf{k},i\omega_n)G^{f}(\textbf{q}+\textbf{k},i\Omega_n+i\omega_{n}), \label{Pi_f}\\
\Pi_c(\textbf{q}, i\Omega_n)&=&\int_{k} G^{c}(\textbf{k},i\omega_{n})G^{c}(\textbf{q}+\textbf{k},i\Omega_n+i\omega_n),
\eea
where $\textbf k$ and $\omega_n$ denote momentum and Matsubara frequency, $G^i$, $i=c,f$ refers to the Green's function of $c$  and $f$ fermion, respectively, and $\int_k \equiv \int_{k_0} \int_{\bf k}$, $\int_{k_0} \equiv T\sum_{\omega_n}$, $\int_{\bf k} \equiv \int \frac{d^2 \bf k}{(2\pi)^2} $. It is easy to check from the saddle point equations that local critical $f$ fermion propagator, i.e., $G^f(\textbf{k}, i \omega_n) = G^f(i \omega_n)$, is always a consistent solution to the saddle point equations. Indeed, at the $M/N \rightarrow 0$ limit, the $f$ fermion propagator is~\cite{Sachdev2015}
\bea\label{Gf}
G^{f}(\tau)=-\frac{\pi^{\frac{1}{4}}\cosh^{\frac{1}{4}}(2\pi \mathcal{E})}{J^{\frac{1}{2}}\sqrt{1+e^{-4\pi \mathcal{E}}}}\bigg(\frac{T}{\sin(\pi T \tau)}\bigg)^{\frac{1}{2}}e^{-2\pi \mathcal{E}T\tau },
\eea
where $\mathcal{E}$ is a parameter controlling the particle-hole asymmetry, and $ \tau \in[0, \beta]$ is the imaginary time. For finite $M/N$, a local critical form of $f$ fermion propagator is still consistent with the full saddle point equations. Moreover, according to Ref.~\cite{Patel2018, Senthil2018}, finite $M/N$ correction is subleading. Thus, we assume the local critical solution holds at a small but finite $M/N$, and focus on the case $M/N \ll 0$.

Moving to the $c$ fermion propagators, we will follow Ref.~\cite{Patel2018} closely. The self-energy of $c$ fermion is given by Eq.~\ref{Sigma_cf}. Since $G^{f}$ is local critical, we can see from Eqs.~(\ref{Sigma_cf}) and~(\ref{Pi_f}) that $\Sigma_{cf}$ is also independent of momentum, i.e., $ \Sigma_{cf}(\textbf{k},i\omega_n) = \Sigma_{cf}(i \omega_n)$, and consequently $\Sigma_{cf}(\tau) = -g^2 G^c(\tau) G^f(\tau) G^f(-\tau)$, with $G^c(\tau) \equiv T \sum_{\omega_n} G^c(i \omega_n)$ and $G^c(i \omega_n) \equiv \int_{\bf k}G^{c}(\textbf{k},i\omega_n)$. Then with the assumption ${\rm sgn(Im}[\Sigma_{cf}(i\omega_n)])=-{\rm sgn}(\omega_n)$, and in the limit of infinite bandwidth $\Lambda \rightarrow \infty$ (i.e., bandwidth is the largest energy scale),
$ G^c(i \omega_n) \approx\nu(0)\int_{-\infty}^{+\infty}\frac{d \epsilon}{2\pi}\frac{1}{i\omega_n-\epsilon-\Sigma_{cf}(\textbf{k},i\omega_n)}=-\frac{i}{2}\nu(0) {\rm sgn}(\omega_n)$, and $G^{c}(\tau)=-\frac{\nu(0)T}{2\sin(\pi T \tau)}$, where $\nu(0)$ is the density of state at fermi level. The self-energy of the $c$ fermion yields~\cite{Patel2018}
\begin{eqnarray}
\Sigma_{cf}^{\rm MFL}(i\omega_n)&=&\frac{ig^2 T}{2J t \cosh^{1/2}(2\pi \mathcal{E})\pi^{3/2}} \notag\\
&&\bigg(\frac{\omega_n}{T}\ln\bigg(\frac{2\pi T e^{\gamma_{E}-1}}{J}\bigg)\notag\\
&&+\frac{\omega_n}{T}\psi\left(-\frac{i \omega_n }{2 \pi  T}\right)+\pi\bigg),
\end{eqnarray}
where $\psi$ is the digamma function, and $\gamma_E=0.577$ is the Euler-Mascheroni constant. The self-energy indicate that in the large bandwidth limit, the $c$ fermions exhibit a MFL behavior.

On the other hand, in the limit where $|i \omega_n + \mu_c - \Sigma^c(i \omega_n)| \gg \Lambda$, one can find local critical solutions of SYK type for both $c$ and $f$ fermions~\cite{Patel2018} at conformal limit. Namely, the $f$ fermion propagator is still given by Eq.~(\ref{Gf}), while the $c$ fermion will enter the IM regime, whose propagator reads~\cite{Patel2018}
\begin{eqnarray}\label{Gc2}
G^c (i\omega_n) \approx\frac{1}{2\pi(\mu_c-\Sigma_{cf}(i\omega_n))},
\end{eqnarray}
where the self-energy is given by
\begin{eqnarray}
\Sigma_{cf}(i\omega_n)&=&\frac{iT^{\frac{1}{2}}g^2 \Lambda^{\frac{1}{2}} \nu^{\frac{1}{2}}(0)(-1)^{\frac{1}{4}}(1+e^{4\pi \mathcal{E}_c })^{\frac{1}{2}}e^{2\pi\mathcal{E}}}{\pi^{\frac{1}{4}}J^{\frac{1}{2}}2^{\frac{3}{2}}(i+e^{2\pi\mathcal{E}_c})\cosh^{\frac{1}{4}}(2\pi\mathcal{E})}\notag\\
 &&\frac{\Gamma(\frac{3}{4}+i\mathcal{E}_c+\frac{\omega_n}{2\pi T})}{\Gamma(\frac{1}{4}+i\mathcal{E}_c+\frac{\omega_n}{2\pi T})},
\end{eqnarray}
where $ \mathcal{E}\simeq -\frac{\mu/J}{\pi^{1/4}\sqrt{2}}$ and $\mathcal{E}_c\simeq -\frac{\pi^{1/4}\cosh^{1/4}(2\pi \mathcal{E})\mu_c}{g}$ at small $\mu_f/J$, $\mu_c/g$ limit. Note Eq.~(\ref{Gc2}) is only valid provided $T\gg T_{\rm inc}$ and $g^2\gg\Lambda J$.

\section{APPENDIX B: SYMMETRY AND NOETHER CURRENTS}
\renewcommand{\theequation}{B\arabic{equation}}
\setcounter{equation}{0}
\renewcommand{\thefigure}{B\arabic{figure}}
\setcounter{figure}{0}
\renewcommand{\thetable}{B\arabic{table}}
\setcounter{table}{0}

In the following, we use Lagrangian formalism to define the energy-momentum tensor in the long wavelength limit. The Lagrangian density of our model is given by
\bea
\mathcal L &=& \sum_l c^\dag_l(x) (\partial_\tau - \frac{\nabla^2}{2m} - \mu_c ) c_l(x) 
\notag\\
&&+ \sum_n f^\dag_n(x) (\partial_\tau - \mu_f ) f_n(x) \notag\\
&& +\sum_{i,j,k,l} \frac{g_{ijkl}}{NM^{1/2}} f_{i}^{\dagger}(x) f_{j}(x) c_{k}^{\dagger}(x) c_{l}(x) \notag\\
&&+\sum_{i,j,k,l} \frac{J_{ijkl}}{N^{3/2}}f_{i}(x)^{\dagger} f_{j}(x)^{\dagger}f_{k}(x)f_{l}(x),
\eea
The Lagrangian $\mathcal L$ is invariant under the translational symmetry $\vec{x}\rightarrow \vec{x}+\vec{a}$, $\tau\rightarrow \tau+a_0$.
 Following the standard Noether procedure,  one obtains  the energy-momentum tensor
\bea
	T_{\mu\nu} = \frac{\partial \mathcal L}{\partial \partial^\mu \psi} \partial_\nu \psi - \delta_{\mu\nu} \mathcal L,
\eea
from which we can get the momentum operator $\mathcal{P}_i \equiv T_{0i} $
\bea
	\mathcal{P}_i(x) = -i\sum_l c^\dag_l(x) \partial_i c_l(x),
\eea
which is conserved due to the translational symmetry. Note that the immobile $f$-fermions do not contribute to the total momentum.  More importantly, the stress tensor used to evaluate the shear viscosity is given by
\bea
	T_{xy}(x) = - \frac1m \sum_l c_l^\dag(x) \partial_x \partial_y c_l(x).
\eea
Indeed, the interacting part of the Lagrangian density does not show up in the stress tensor.  Only the diagonal part is modified,
\bea
	T_{xx}(x) &=& - \frac1m \sum_l c_l^\dag(x) \partial_x^2 c_l(x) - \mathcal L(x) \\
	&=&  -\sum_l c_l^\dag(x)\Big(\partial_\tau - \frac{-\partial_x^2+ \partial_y^2}{2m}-\mu_c \Big) c_l(x) \notag\\
	&&- \sum_n f^\dag_n(x) (\partial_\tau - \mu_f ) f_n(x)- \mathcal L_I,
\eea
where the last term is the interacting part.

\section{APPENDIX C: THE DERIVATION OF SHEAR VISCOSITY IN TERMS OF SPECTRAL FUNDTION}
\renewcommand{\theequation}{C\arabic{equation}}
\setcounter{equation}{0}
\renewcommand{\thefigure}{C\arabic{figure}}
\setcounter{figure}{0}
\renewcommand{\thetable}{C\arabic{table}}
\setcounter{table}{0}

We prove that the shear viscosity defined via the Kubo formula
\begin{eqnarray}
\eta &=&\lim_{\omega\rightarrow 0}\frac{1}{\omega}{\rm Im} G^{R}_{xy,xy}(\omega,0),\nonumber\\
G^{R}_{xy,xy}(\omega,0)&=&-i\int dt d\vec{x}e^{i\omega t}\theta(t)\notag\\
&&\langle[T_{xy}(t,\vec{x}),T_{xy}(0,0)]\rangle,
\end{eqnarray}
is equivalent to (\ref{eta}) in terms of spectral functions.

The $xy$-component of the uniform energy-momentum tensor for $c$-fermions is given by
\be
T_{xy}=\int \frac{d^2 \bf k}{(2\pi)^2} c^{\dagger}_{ki} \frac{k_x k_y}{m}c_{ki}.
\ee
To obtain the retarded Green function, we first use the imaginary time formula. In the tree level, we have
\bea
G_{xy,xy}(i\Omega,0)&=&-MT\sum_{\omega_n} \int \frac{d^2 \bf k}{(2\pi)^2} \bigg(\frac{k_xk_y}{m}\bigg)^2 \nonumber\\
&& G_{c}(i\omega_n,k)G_{c}(i\omega_n+i\Omega_n,k).
\eea
Using the spectral representation, $G(z)=\int \frac{d \omega}{2\pi} \frac{A^c(\omega)}{z-\omega}$, one is able to sum over Matsubara frequencies and continue to real frequency
\bea
&&{\rm Im} T\sum_{\omega_n}G(i\omega_n)G(i\omega_n+\Omega+i\delta)\notag\\
&=&-\frac{1}{2}\int \frac{d\omega'}{2\pi}A^c(\omega')A^c(\omega'+\Omega)\notag\\
&&[n_{F}(\omega')-n_{F}(\omega'+\Omega)].
\eea
We obtain the imaginary part of the retarded Green's function
\bea
&&{\rm Im} G^{R}_{xy,xy}(\Omega,0)\nonumber\\
&=&\frac M2 \int \frac{d^2 \bf k}{(2\pi)^2} \bigg(\frac{k_x k_y}{m}\bigg)^2 
\int \frac{d\omega}{2\pi} A^c(\omega,k)A^c(\omega+\Omega,k)\notag\\
&&[n_{F}(\omega)-n_{F}(\omega+i \Omega)].
\eea
The shear viscosity is then given by
\bea
\eta=\frac M2\int_{-\infty}^\infty \frac{d \omega}{2\pi}\bigg(-\frac{\partial n_{F}}{\partial \omega}\bigg) \int_{-\infty}^{\infty} d\epsilon \Theta_{xy}(\epsilon) A^c(\omega,\epsilon)^2,
\eea
where $ \Theta_{xy}(\epsilon) \equiv \int \frac{d^2 \bf k}{(2\pi)^2} \big(\frac{k_x k_y}{m}\big)^2 \delta(\epsilon-\epsilon_k) $.

\section{APPENDIX D: SHEAR VISCOSITY IN MARGINAL FERMI LIQUID}
\renewcommand{\theequation}{D\arabic{equation}}
\setcounter{equation}{0}
\renewcommand{\thefigure}{D\arabic{figure}}
\setcounter{figure}{0}
\renewcommand{\thetable}{D\arabic{table}}
\setcounter{table}{0}

In MFL regime, the well-defined fermi surface allows us to approximate the  density of states $\nu(\epsilon)$ at energy $\epsilon $ by density of states at fermi surface $\nu(0)$. Then we have
\bea
 \Theta_{xy}(\epsilon) &=& m^2 v_F^4 \int \frac{d^2 \bf k}{(2\pi)^2}  \cos^2 \theta \sin^2 \theta  \delta(\epsilon-\epsilon_k) \notag\\
 &\approx& \frac{ m^2 v_F^4}{16 \pi } \nu(0) \approx \frac{\nu(0)}{16 \pi m^2} ,
\eea
where in the last step, we use the relation $v_F \sim \frac1m$ in the isotropic dispersion. The shear viscosity is given by
\bea
	 \eta_{\rm MFL} &=& \frac{M }{16 \pi T}  \int d\omega \sech^2(\frac{\omega}{2T}) \int d \epsilon \Theta_{xy}(\epsilon)A_{\rm MFL}^c(\omega, \epsilon)^2 \notag\\
	 &=& \frac{M}{16 \pi T} \frac{m^2 v_F^4}{16 \pi} \nu(0) \int d\omega \sech^2(\frac{\omega}{2T}) \int d\epsilon A_{\rm MFL}^c(\omega, \epsilon)^2 \notag\\
	 &=& \frac{Mm^2 v_F^4 \nu(0)}{128 \pi T}  \int d\omega  \frac{\sech^2(\frac{\omega}{2T})}{|\Im \Sigma^{\rm MFL}_{cf}(\omega)|} \notag\\
	 &\approx& 0.0300627\frac{Mt^2J}{g^2T}\cosh^{\frac12}(2\pi\mathcal{E}),
\eea
where in the last line, we have used the relation $v_F \sim \frac1m \sim \frac1{\nu(0)} \sim t$ in the isotropic dispersion.

\section{APPENDIX E: SHEAR VISCOSITY IN INCOHERENT METAL}
\renewcommand{\theequation}{E\arabic{equation}}
\setcounter{equation}{0}
\renewcommand{\thefigure}{E\arabic{figure}}
\setcounter{figure}{0}
\renewcommand{\thetable}{E\arabic{table}}
\setcounter{table}{0}

For the dispersion relation $\epsilon_k=\frac{k^2}{2m}-\frac \Lambda2$ with bandwidth $\epsilon_k \in [-\frac\Lambda2,\frac\Lambda2]$, we have
\bea
	\Theta_{xy}(\epsilon)
	&=&\int \frac{d^2 \bf k}{(2\pi)^2} \bigg(\frac{k_x k_y}{m}\bigg)^2 \delta(\epsilon-\epsilon_k)	\notag\\
	&=&\frac{1}{(2\pi m)^2}\int d\theta \cos^2\theta\sin^2\theta \int dk k^5 \delta(\epsilon-\epsilon_k) \notag\\
	&=& \frac m{4\pi} \left(\epsilon+\frac\Lambda2 \right)^2 \theta\left(\frac\Lambda2-|\epsilon|\right),
\eea
where $ \theta(x)$ is the unit step function. One can also find $\Theta_{xy}$ using Fourier transform~\cite{Pakhira2015, Tremblay2013S}, which exactly gives the same result. The spectral function of $c$ fermion in IM region is given by~\cite{Patel2018},
\bea
&&A^c(\omega, \epsilon)= A^c(\omega)\notag\\
&=&-2\Re \Big[ \frac{e^{i \frac{3\pi}4}\pi^{1/4} J^{1/2} \cosh^{1/4}(2\pi \mathcal{E}) (i + e^{2\pi \mathcal{E}_c})}{g T^{1/2} \sqrt{1+e^{4\pi \mathcal{E}_c}}} \notag\\
&&\frac{\Gamma( \frac14- i \frac{\beta \omega - 2\pi \mathcal{E}_c}{2\pi})}{\Gamma (\frac34- i \frac{\beta \omega - 2\pi \mathcal{E}_c}{2\pi} )}\Big],
\eea
which is independent of $\epsilon$ as a result of local criticality. Then the shear viscosity is given by
\bea
\eta&=& \frac{M }{16 \pi T} \int d \epsilon \Theta_{xy}(\epsilon) \int d\omega \sech^2(\frac{\beta \omega}{2}) A^c(\omega)^2  \notag\\
&=& \frac{M }{16 \pi T} \frac{\Lambda^2}{12\pi} \frac{8 \pi^{5/2} J \cosh^{1/2}(2\pi \mathcal{E})}{g^2  \cosh(2\pi \mathcal{E}_c)}
\notag\\
 &=& \frac{M \pi^{1/2}}{24} \frac{\Lambda^2 J}{g^2 T}  \frac{\cosh^{1/2}(2\pi \mathcal{E})}{\cosh(2\pi \mathcal{E}_c)},
\eea
where we have used $\int d \epsilon \Theta_{xy}(\epsilon) = \frac{\Lambda^2}{12\pi}$, and
\bea
	&&\int d\omega \sech^2(\frac{\beta \omega}{2}) A^c(\omega)^2 \notag\\
	&=& \frac{16 \pi^{5/2} J \cosh^{1/2}(2\pi \mathcal{E})}{g^2 T} \frac{1}{2\cosh(2\pi \mathcal{E}_c)} \notag\\
	&&\int d\omega \Big( \frac{\sech(\beta \omega-2\pi \mathcal{E}_c)}{\Gamma (\frac34+ i \frac{\beta \omega - 2\pi \mathcal{E}_c}{2\pi} )\Gamma (\frac34- i \frac{\beta \omega - 2\pi \mathcal{E}_c}{2\pi} )} \Big)^2 \notag\\
	&=& \frac{8 \pi^{5/2} J \cosh^{1/2}(2\pi \mathcal{E})}{g^2  \cosh(2\pi \mathcal{E}_c)} \int dx \Big( \frac{\sech(x)}{\Gamma (\frac34+ i \frac{x}{2\pi} )\Gamma (\frac34- i \frac{x}{2\pi} )} \Big)^2 \notag\\
	&=& \frac{8 \pi^{5/2} J \cosh^{1/2}(2\pi \mathcal{E})}{g^2  \cosh(2\pi \mathcal{E}_c)}.
\eea

\section{APPENDIX F: THERMAL DIFFUSION CONSTANT}
\renewcommand{\theequation}{F\arabic{equation}}
\setcounter{equation}{0}
\renewcommand{\thefigure}{F\arabic{figure}}
\setcounter{figure}{0}
\renewcommand{\thetable}{F\arabic{table}}
\setcounter{table}{0}

We calculate the thermal diffusion coefficient in both regimes by using the results given in Ref.~\cite{Patel2018}.
The thermal diffusivity can be given by Einstein's relation
\begin{equation}\label{dif}
\mathcal{D}=\frac{\kappa_0}{c_V},
\end{equation}
where $\kappa_0$ is the `closed-circuit' thermal conductivity and $c_V$ is the specific heat.

In MFL regime, from Ref.~\cite{Patel2018}, we have $\kappa_0^{MFL}\sim MJt^2/g^2$ and $c^{MFL}_V\sim M (g^2/t^2)(T/J) \ln(J/T)$, where we have set $\mathcal{E}=0$ in the following calculations. The thermal diffusion constant scales as
\begin{equation}
\mathcal{D}^{MFL}\sim \frac{ J^2 t^4}{g^4 T  \ln (\frac{J}{T})}.
\end{equation}
Note that as $T\rightarrow 0$, the thermal diffusion constant becomes divergent same as the shear viscosity. Since $T\ll T_{inc}$, we conclude that
$
\mathcal{D}_{\kappa}^{MFL}\gg \frac{t^2}{g^2}J\frac{1}{\ln (g/t)}.
$

Similarly, in the IM regime, one has $\kappa_0^{IM}\sim MJ\Lambda^2/g^2$ and $c^{IM}_V\sim MJT/g^2$
Ref.~\cite{Patel2018}.
The thermal diffusion constant scales as
\begin{equation}
\mathcal{D}^{IM}\sim\frac{\pi ^{5/2} \Lambda ^2}{64 T}.
\end{equation}
Due to the IM existing only at temperature above $T_{inc}$, we always have $\mathcal{D}^{IM}\ll \frac{\pi ^{5/2} g^2}{64 J}$. 
In the MFL regime, the thermal diffusion has a $1/T$ dependence due to local criticality.  It was argued that the fast `Planckian' dissipation together with the causality of diffusion results in an upper bound of diffusivity~\cite{hartman}. The results found in this work strongly implies that the shear viscosity and the upper bound of diffusivity maybe deeply connected.

\section{APPENDIX G: RELATION TO THE DC CONDUCTIVITY}
\renewcommand{\theequation}{G\arabic{equation}}
\setcounter{equation}{0}
\renewcommand{\thefigure}{G\arabic{figure}}
\setcounter{figure}{0}
\renewcommand{\thetable}{G\arabic{table}}
\setcounter{table}{0}

In MFL regime, similar to case of shear viscosity, the inverse lifetime Eq.~(9) also gives rise to the $T^{-1}$  dependence of DC conductivity. From \cite{Patel2018}, one has
\begin{equation}
	\sigma_{DC}^{MFL}\sim  \frac{M}{m \gamma} \sim \frac{MJt^2}{Tg^2}.
\end{equation}
From uncertainty principle, the metallic conductivity in 2D is bounded below by the Mott-Ioffe-Regel (MIR) limit, and $\sigma=n \tau /m \sim (k_F l) 1/\hbar\geq 1/\hbar$, where $l$ is the electronic mean free path and the charge unit is omitted. The conductivity obtained here can be lower than the MIR limit $1/\hbar$ numerically by tuning parameters, although the MFL is not rigorously a bad metal.

In the IM regime, the DC conductivity reads
\begin{equation}\label{sigmaIM}
\sigma^{\rm IM}_{\rm DC}\sim \frac{M \Lambda^2 J}{g^2 T}\frac{\cosh^{1/2}(2\pi \mathcal{E})}{\cosh(2\pi \mathcal{E}_c)},
\end{equation}
which shares the same scaling form with the shear viscosity in Eq.~(10). It is not surprising. Firstly, because of local criticality, the spectral density is independent of momentum. Secondly, the vertex of shear viscosity and conductivity has the same scaling, which is $1/m\sim t$. The combination of above two features completely determine the scaling form.

Both of shear viscosity and DC conductivity vanish when $T\gg T_{inc}$ due to the same scaling forms in Eqs.~(10) and (\ref{sigmaIM}). To reach $T\gg T_{inc}$, one can consider the decouple limit $t\to0$ while keeping other couplings and temperature fixed, which agrees with the fact that transport coefficients die out. Furthermore, the entropy $\mathcal{S}_c$ contributed by c-fermion keep fixed under the decouple limit, which is equal to the entropy of the SYK model with $J_{IM}=g^2/J$.
From this point of view, the violation of the KSS bound of $\eta/\mathcal{S}_c$ here shares the same reason with the deviation from the MIR limit of $\sigma_{DC}$ in the incoherent metal regime.

These two bounds can be understood from the inverse lifetime for the $c$ fermions Eq.~(9). In the MFL regime with temperature $T\ll T_{inc}$ and $t\gg g$, $J\gg T$, the $c$ fermions' lifetime behaves as $\tau_{h}\sim T_{inc}/(tT)\gg 1/t$. However, in the IM regime, due to local criticality, the universal `Planckian' time $\tau_h\sim 1/T$ give the temperature dependence of  transport coefficients.


\begin{thebibliography}{99}
\bibitem{note1} In classical limit, the shear viscosity is given by $\eta = \frac13 n p l_{\text{mfp}}$ where $n$ is the density, the mean free path is $l_{\text{mfp}} \propto T$, and the average momentum is $p \propto T^{1/2}$. On the other hand, the entropy density tends to a constant set by the local Hilbert space. In conclusion, $\eta/\mathcal{S} \propto T^{3/2}$.
\bibitem{notex}For an electron gas in metals, there is a  minimum metallic conductivity $\sigma=\frac{n \tau e^2}{m} \sim (k_F l)k^{d-2}_F \frac{e^2}{\hbar}\geq k^{d-2}_F \frac{e^2}{\hbar}$, where  $k_F$ is the Fermi velocity, and $l$ is the electronic mean free path, respectively.
\end{thebibliography}
\end{document}